# Electronic structure of Ti-doped $Sr_4Sc_2Fe_2As_2O_6$

I. R. Shein* and A. L. Ivanovskii

*Institute of Solid State Chemistry, Ural Division, Russian Academy of Sciences, Pervomaiskaya St., 91, Yekaterinburg, 620990 Russia*

**Abstract**

**First principle FLAPW-GGA calculations have been performed with the purpose to understand the electronic properties for the newly synthesized tetragonal (space group *P4/nmm*) layered iron arsenide-oxide: $Sr_4Sc_2Fe_2As_2O_6$ doped with titanium. The total and partial densities of states, low-temperature electron specific heat and molar Pauli paramagnetic susceptibility have been obtained and discussed for $Sr_4ScTiFe_2As_2O_6$ in comparison with parent phase $Sr_4Sc_2Fe_2As_2O_6$.**

**Our results show that the insertion of Ti into Sc sublattice of iron arsenide-oxide phase $Sr_4Sc_2Fe_2As_2O_6$ leads to principal change of its electronic structure; in result the insulating so-called "charge reservoirs', *i.e.* perovskite-based [$Sr_4Sc_2O_6$] blocks became conducting. This situation differs essentially from the known picture for all others Fe-As superconductors where the conducting [$Fe_2As_2$] blocks are separated by isolating blocks.**

* Corresponding author.
*E-mail address:* shein@ihim.uran.ru (I.R. Shein).

The recent discovery of superconductivity at $T_c$'s up to 55-56 K in the layered iron-pnictide systems (reviews [1-2]) has led to an intensive search for related superconductors (SCs). Very recently, in addition to the known groups of such materials (so-called "1111", "122" and "111" phases and some others, see [1,2]) a set of novel more complex iron-based compounds with different structures have been found. Among them are very interesting five-component phases such as $Sr_3Sc_2Fe_2As_2O_5$ [3], $Sr_4Sc_2Fe_2P_2O_6$ [4], $Sr_4Sc_2Fe_2As_2O_6$, $Sr_4Cr_2Fe_2As_2O_6$ [5,6] and $Sr_4Sc_2Co_2As_2O_6$ [6].

These materials are isostructural and adopt tetragonal (space group *P4/nmm*) crystal structure with alternating stacking of [$TM_2Pn_2$] (*TM* – transition metals, *Pn* - pnictogens) and perovskite-based [$Sr_4Sc_2O_6$] blocks, where [$TM_2Pn_2$]/[$TM_2Pn_2$] separation ($l \geq 15.5$ Å) is the longest among all Fe-based SCs [1,2]. The available data [3-6] demonstrate that iron phosphide-oxide phase $Sr_4Sc_2Fe_2P_2O_6$ is superconductor with critical temperature $T_c$ ~17K, whereas others above As-containing materials are non-superconducting. It can be caused as by their structural features [5], or by deficiency of carrier concentration [5,6].

According to the preliminary data [7] the partial substitution of Sc by Ti in $Sr_4Sc_2Fe_2As_2O_6$ (as well as in $Sr_3Sc_2Fe_2As_2O_5$) which results in an increased carrier concentration, leads to the appearance of superconductivity. The authors [7] assert, that $Sr_4Sc_2Fe_2As_2O_6$ superconducts when it is doped with 30% Ti on Sc sites; besides, the onset of superconductivity occurs at about 45K and full bulk superconductivity is observed until 7K for the sample of $Sr_4Sc_{0.8}Ti_{1.2}Fe_2As_2O_6$.

In view of these circumstances, in this Communications we present a preliminary data of our studying the effect of Ti doping on the electronic properties of $Sr_4Sc_2Fe_2As_2O_6$ and we focus our attention on the variation of the density of states (DOS) as result from the Ti insertion into Sc sublattice.

Our calculations were carried out by means of the full-potential method with mixed basis APW+lo (FLAPW) implemented in the WIEN2k suite of programs [8]. The generalized gradient approximation (GGA) to exchange-correlation potential in the PBE form [9] was used. The plane-wave expansion was taken up to $R_{MT} \times K_{MAX}$ equal to 7, and the $k$ sampling with 13×13×5 $k$-points in the Brillouin zone was used. At the first stage the full-lattice optimization for parent phase $Sr_4Sc_2Fe_2As_2O_6$ including the atomic positions was done. The self-consistent calculations were considered to be converged when the difference in the total energy of the crystal did not exceed 0.1 mRy and the difference in the total electronic charge did not exceed 0.001 $e$ as calculated at consecutive steps. The data are presented in Table 1 and are in reasonable agreement with available experiments [5-7].

For Ti-doped iron arsenide-oxide $Sr_4Sc_2Fe_2As_2O_6$ the calculated lattice constants are used; and the Ti-containing phase with formal stoichiometry $Sr_4ScTiFe_2As_2O_6$ was simulated by the partial replacement Sc → Ti as depicted in Figure 1.

Figure 2 shows the total and partial DOSs of $Sr_4Sc_2Fe_2As_2O_6$ in comparison with $Sr_4ScTiFe_2As_2O_6$. For parent phase $Sr_4Sc_2Fe_2As_2O_6$ the occupy bands form three main groups (A-C) in the intervals -12.0 eV ÷ -10.8 eV (peak A); -5.3 ÷ -2.1 eV (peak B) and from – 2.1 eV up to $E_F$ (peak C). The Fermi level is crossed by low-dispersive bands with mainly Fe 3$d$ character; these bands form two electron pockets centered at $M$ and three hole pockets centered at Γ; for details see [10]. For $Sr_4Sc_2Fe_2As_2O_6$ the DOS peak A is almost completely made from As 4$s$ orbitals - with very small admixture of Sr 5$p$ orbitals. The contributions from the valence s,$p$ states of Sr to the all occupied bands are very small, *i.e.* in $Sr_4Sc_2Fe_2As_2O_6$ these atoms are in the form of cations close to $Sr^{2+}$.

In the next interval from -5.3 eV up to -2.1 eV (peak B) all the atoms of $Sr_4Sc_2Fe_2As_2O_6$ phase are contributing to the DOS and these states are responsible for the hybridization effects, i.e. for inter-atomic covalent bonding. Taking into account the distribution of the corresponding atoms over the [$Fe_2As_2$] and [$Sr_4Sc_2O_6$] blocks (see Fig. 1) it points to the formation of the Fe–As and Sc–O covalent bonds due to the hybridization of Fe 3*d* - As 4*p* states and Sc 3*d* - O 2*p* states, respectively, see also below. The low-ling empty states (peak D) are mainly Fe 3*d* character.

Let us tote that the occupied states near the Fermi level in $Sr_4Sc_2Fe_2As_2O_6$ (peak C) are formed ***exclusively*** by states of [$Fe_2As_2$] blocks, whereas perovskite-like [$Sr_4Sc_2O_6$] blocks are insulating. Therefore, the conduction in this phase (as well as for others known families of "1111", "122" and "111" FeAs SCs [1,2]) is strongly anisotropic, *i.e.* happening in the [$Fe_2As_2$] blocks. Notice that the contributions from As states are much smaller than the contributions from Fe 3*d* orbitals, see Table 2, where the total and orbital decomposed partial DOSs at the Fermi level, $N(E_F)$, are shown.

The electronic structure for $Sr_4ScTiFe_2As_2O_6$ differs sharply, see Figure 2, Table. 2. Firstly, a set of new DOSs peaks; B′ and D′, as well as the forbidden gap at -2 eV appears. Secondly, the width of valence zone increases: from 5.3 eV (for $Sr_4Sc_2Fe_2As_2O_6$) up to 6.9 eV (for $Sr_4ScTiFe_2As_2O_6$).

But the most important difference for $Sr_4ScTiFe_2As_2O_6$ *versus* $Sr_4Sc_2Fe_2As_2O_6$ is the structure of near-Fermi region, where the Ti 3*d* states move in the valence zone region and these states begin to fill. Besides, in this region the admixtures of states of others atoms from [$Sr_4ScTiO_6$] blocks are added. In result, unlike the insulating type of "pure" [$Sr_4Sc_2O_6$] blocks in $Sr_4Sc_2Fe_2As_2O_6$, the Ti-doped [$Sr_4ScTiO_6$] blocks became conducting. Moreover, for $Sr_4ScTiFe_2As_2O_6$ the density of Ti 3*d* states at the Fermi level is about 43% higher, than the density of Fe 3*d* states for [$Fe_2As_2$] blocks, whereas the values of $N^{Fe3d}(E_F)$ and $N^{As4p}(E_F)$ for $Sr_4ScTiFe_2As_2O_6$ are decreased as compared with $Sr_4Sc_2Fe_2As_2O_6$, Table 2. The obtained data allow us to estimate the Sommerfeld constants ($\gamma$) and the Pauli paramagnetic susceptibility ($\chi$) for examined phases under assumption of the free electron model as $\gamma = (\pi^2/3)N(E_F)k^2_B$ and $\chi = \mu_B^2 N(E_F)$. It is seen from Table 2 that both $\gamma$ and $\chi$ increase at replacement of Sc by Ti.

In summary, our results show that the insertion of Ti into Sc sublattice of iron arsenide-oxide phase $Sr_4Sc_2Fe_2As_2O_6$ leads to principal change of its electronic structure; in result the insulating so-called "charge reservoirs', *i.e.* perovskite-based [$Sr_4Sc_2O_6$] blocks became conducting. This situation differs essentially from the known picture for all others Fe-*Pn* superconductors (such as above “1111”, “122” and “111” phases) where the conducting [$Fe_2Pn_2$] blocks are separated by isolating blocks. Therefore the report [7] is seems quite surprising.

**ACKNOWLEDGMENTS**

Financial support from the RFBR (Grant 09-03-00946-a) is gratefully acknowledged.

---

**References**

[1] A.L. Ivanovskii, Physics - Uspekhi **51** 1229 (2008).
[2] M.V. Sadovskii, Physics - Uspekhi **51** 1201 (2008)
[3] X. Zhu, F. Han, G. Mu, B. Zeng, P. Cheng, B. Shen and H. Wen, Phys. Rev.B **79** 024516 (2009).
[4] H. Ogino, Y. Matsumura, Y. Katsura, K. Ushiyama, S. Horii, K. Kishio and J. Shimoyama, *arXiv*:0903.3314 (unpublished)
[5] H. Ogino, Y. Katsura, S. Horii, K. Kishio and J. Shimoyama, *arXiv*:0903.5124 (unpublished)
[6] Y.L. Xie, R.H. Liu, T. Wu, G. Wu, Y.A. Song, D. Tan, X.F. Wang, H. Chen, J.J. Ying, Y.J. Yan, Q.J. Li and X.H. Chen, *arXiv*:0903.5484 (unpublished)
[7] G.F. Chen, T.L. Xia, P. Zheng, J.L. Luo, and N.L. Wang, *arXiv*:0903.5273 (unpublished)

[8] P. Blaha, K. Schwarz, G. K. H. Madsen, D. Kvasnicka, and J. Luitz, WIEN2k, *An Augmented Plane Wave Plus Local Orbitals Program for Calculating Crystal Properties,* (Vienna University of Technology, Vienna, 2001)
[9] J. P. Perdew, S. Burke, and M. Ernzerhof, Phys. Rev. Lett. **77**, 3865 (1996)
[10] I.R. Shein, and A.L. Ivanovskii. *arXiv*:0903.4038 (unpublished)

Table 1. The optimized lattice parameters (*a* and *c*, in Å) and atomic positions for $Sr_4Sc_2Fe_2As_2O_6$ in comparison with experiment [5].

| $Sr_4ScTiO_6Fe_2As_2$ | | | | | | | |
|---|---|---|---|---|---|---|---|
| | | | | $Sr_4Sc_2O_6Fe_2As_2$ | | | |
| *Space group* | *Pmma* | | | *P4/nmm* | | | |
| *a* | 5.708 | | | 4.036 (4.050) | | | |
| *c* | 15.534 | | | 15.534 (15.809) | | | |
| $Sr_1$ (2e) | 0.25 | 0.0 | 0.8176 | $Sr_1$ (2c) | 0.25 | 0.25 | 0.3176 (0.3113) |
| $Sr_2$ (2e) | 0.25 | 0.0 | 0.5870 | $Sr_2$ (2c) | 0.25 | 0.25 | 0.0870 (0.0847) |
| $Sr_3$(2f) | 0.25 | 0.5 | 0.1834 | Sc(2c) | 0.25 | 0.25 | 0.8042 (0.8071) |
| $Sr_4$(2f) | 0.25 | 0.5 | 0.4130 | $O_1$ (4f) | 0.75 | 0.25 | 0.2179 (0.2143) |
| $O_1$ (2e) | 0.25 | 0.0 | 0.4325 | $O_2$ (2c) | 0.25 | 0.25 | 0.9325 (0.9301) |
| $O_2$ (2f) | 0.25 | 0.5 | 0.5675 | Fe (2b) | 0.75 | 0.25 | 0.5 |
| $O_3$ (8l) | 0.0 | 0.75 | 0.7179 | As (2c) | 0.25 | 0.25 | 0.5778 (0.5854) |
| Fe (4g) | 0.0 | 0.75 | 0.0 | | | | |
| $As_1$(2e) | 0.25 | 0.0 | 0.0778 | | | | |
| $As_2$(2f) | 0.25 | 0.5 | 0.9222 | | | | |
| Ti(2f) | 0.25 | 0.5 | 0.6958 | | | | |
| Sc(2e) | 0.25 | 0.0 | 0.3042 | | | | |

* for $Sr_4Sc_2Fe_2As_2O_6$ the experimental data (in parenthesis [5]) are given by displacement at *c*/2.

Table 2. Total and partial densities of states at the Fermi level (in states/eV·form.unit), electronic heat capacity γ (in $mJ \cdot K^{-2} \cdot mol^{-1}$) and molar Pauli paramagnetic susceptibility χ (in $10^{-4}$ emu/mol) for $Sr_4Sc_2Fe_2P_2O_6$ and $Sr_4Sc_2Fe_2As_2O_6$ and $Sr_4ScTiFe_2As_2O_6$.

| system | $N^{Fe3d}(E_F)$ | $N^{Ti3d}(E_F)$ | $N^{Asp}(E_F)$ | $N^{tot}(E_F)$ | γ | χ |
|---|---|---|---|---|---|---|
| $Sr_4Sc_2Fe_2As_2O_6$ | 2.779 | - | 0.081 | 3.754 | 8.85 | 1.21 |
| $Sr_4ScTiFe_2As_2O_6$ | 2.278 | 3.254 | 0.043 | 8.609 | 20.29 | 2.77 |

## FIGURES

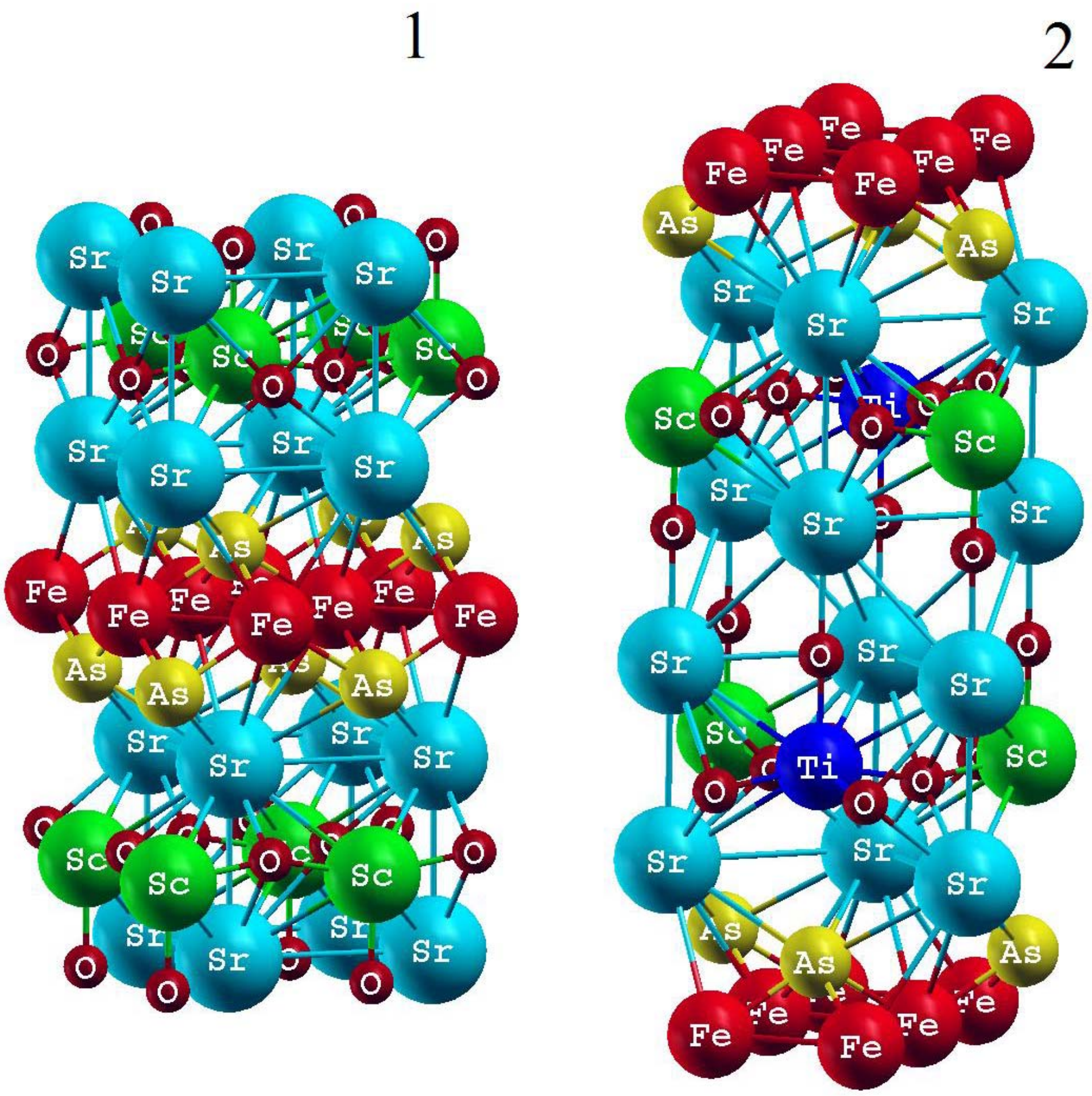

FIG. 1. Crystal structure of the phase $Sr_4Sc_2Fe_2As_2O_6$ (1) and $Sr_4ScTiFe_2As_2O_6$ (2).

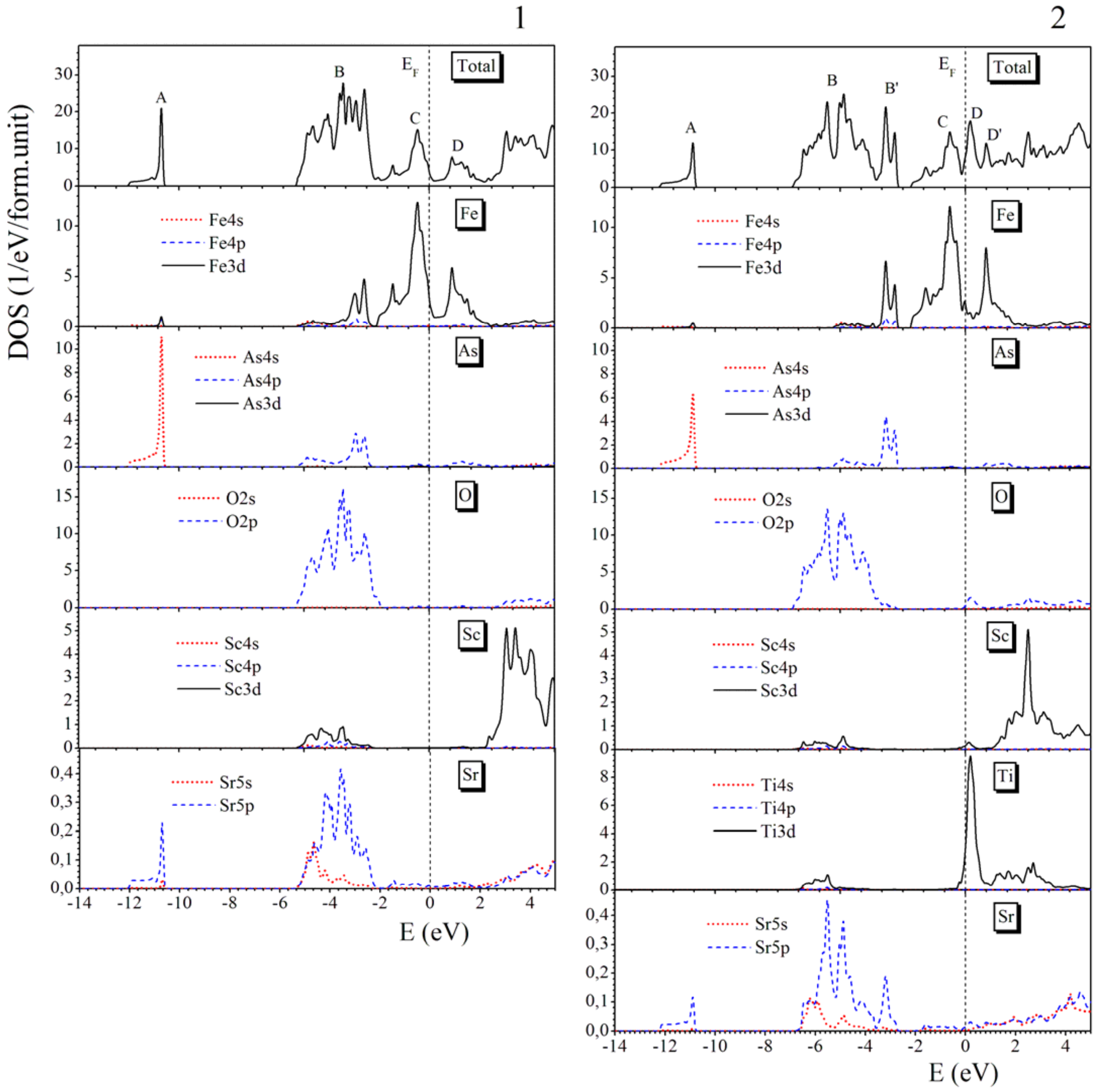

Fig. 2. Total (*upper panels*) and partial densities of states (*bottom panels*) for $Sr_4Sc_2Fe_2As_2O_6$ (1) and $Sr_4ScTiFe_2As_2O_6$ (2).